\documentclass[showpacs,twocolumn,aps,floatfix,superscriptaddress]{revtex4}
\usepackage{amsmath,amssymb,eucal,graphicx,subfigure}

\usepackage{color}

\begin{document}

\def\a{\alpha}
\def\g{\gamma}
\def\av#1{\langle#1\rangle}

\title{Spatially Distributed Social Complex Networks}

\author{Gerald F.~Frasco}\email{frascogf@clarkson.edu}
\affiliation{Department of Physics, Clarkson University, Potsdam, NY
13699-5820}

\author{Jie Sun}\email{sunj@clarkson.edu}
\affiliation{Department of Mathematics \& Computer Science, Clarkson University, Potsdam, NY 13699-5815}

\author{Hern\'an D.~Rozenfeld}\email{hernanrozenfeld@gmail.com}
\affiliation{Editorial Office, American Physical Society, 1 Research Road, Ridge, NY 11961}

\author{Daniel ben-Avraham}\email{benavraham@clarkson.edu}
\affiliation{Department of Physics, Clarkson University, Potsdam, NY
13699-5820} 
\affiliation{Department of Mathematics \& Computer Science, Clarkson University, Potsdam, NY 13699-5815}

\begin{abstract}
We propose a bare-bones stochastic model that takes into account {\it both} the geographical distribution of people within a country and their complex network of connections. The model, which is designed to give rise to a scale-free network of social connections and to visually resemble the geographical spread seen in satellite pictures of the Earth at night, gives rise to a power-law distribution for the ranking of cities by population size (but for the largest cities) and reflects the notion that highly connected individuals tend to live in highly populated areas. It also yields some interesting insights regarding Gibrat's law for the rates of city growth (by population size), in partial support of the findings in a recent analysis of real data [Rozenfeld {\it et al.}, Proc.~Natl.~Acad.~Sci.~U.S.A {\bf105}, 18702 (2008)]. The model produces a nontrivial relation between city population and city population density and a superlinear relationship between social connectivity and city population, both of which seem quite in line with real data.
\end{abstract}

\pacs{%
89.75.Hc,  
02.50.-r   
}
\maketitle

\section{Introduction}
The precise nature of the geographical distribution of human populations and their concurrent network of social contacts has drawn considerable interest in recent years because of its critical role in the spread of epidemics and in developing effective immunization strategies for their arrest~\cite{Colizza,Castellano,Boguna,Belik}, and its effect on the evolution of the electoral map during election times, on the spread of rumors and ideas~\cite{Takaguchi}, and on commerce, transportation, and city planning~\cite{Krugman,Eeckhout,Batty,Bettencourt}.

One way to address the issues of geographical distribution of populations is by studying data that are available through census bureaus~\cite{Census,Eeckhout}, which provide detailed information on the distribution of the population in a given region, usually in 10-year snapshots (as in the USA). In addition, there have been other proposals that explain or reflect the global distribution of population and the geographical shape of cities based on models of diffusion-limited aggregation~\cite{Makse}, and on the idea of ``demographic gravitation"~\cite{barabasi} and subsequent developments~\cite{Rybski}. On the other hand, there is a wealth of literature on the structure of large-scale complex networks through the exploration of massive data sets, which has led to the discoveries of the ``small-world"~\cite{Watts98} and ``scale-free"~\cite{Barabasi99} properties as key ingredients of such networks.  A most recent study addresses the relation between social connectivity and city population, exploiting country-scale mobile communication data~\cite{Schlapfer2012}.

The problem of how social networks are embedded in space (or alternatively, how geographically distributed populations are socially connected) is poorly understood from a physical point of view. A possible approach is to build a computer simulation of the population in question, as faithful to reality as attainable by one's resources. While this approach addresses real practical questions in an obvious, direct way, it perforce  relies on a wealth of parameters that must be  tweaked and tuned individually to mirror reality.  This wealth of parameters makes such models unwieldy from a theoretical point of view and makes it hard to tease out the precise underlying causes for a particular kind of observed behavior, even in purely numerical studies.  Here, in contrast, our goal is to introduce a minimalistic kind of model---one that reproduces the salient features of real-life populations with as few parameters as possible---sacrificing detail for the sake of a better shot at physical insight and a deeper understanding of cause and effect.

Our model, presented in Sec.~\ref{model.sec}, merges two powerful ideas, selected for their analytical simplicity and their success in capturing key features of human populations: The network of contacts is generated by a variant of the Krapivsky-Redner (KR) network growth model~\cite{KR}, while the placement of the nodes in space is based on Kleinberg's observation that navigation of a small-world network is {\it optimal} when the long-range contacts follow a {\it specific} power-law distribution~\cite{kleinberg}.  Combining these key elements, the model reproduces a scale-free social network of contacts, as well as a spatial distribution of city sizes that resembles reality---as compared with satellite views of the Earth at night---while using only a dearth of parameters. 

A rudimentary analysis and the model's results are presented in Sec.~\ref{results.sec}, where we show that our ``city" populations exhibit a power-law Zipf-like rank-size distribution for all but the largest cities. We further analyze the rate of growth of cities, finding support for the  results of Rozenfeld {\it et al.}~\cite{hernan} that the rate of growth and its standard deviation in real-life data decline with city population, in contradistinction with Gibrat's law (that these quantities are constant). The population density of cities grows as a nontrivial function of their populations, a result that also seems consistent with real data. Finally, we find that our model yields a superlinear correlation between social connectivity and city population, supporting the novel finding in Ref.~\cite{Schlapfer2012}. Our model also produces a positive correlation between highly connected individuals and highly populated areas, a plausible notion that has not yet been quantified in real data. We conclude in Sec.~\ref{discussion.sec} with a discussion of our findings, interesting open problems, and ideas for further work.

\section{Model Description}
\label{model.sec}
Our model describes how to place a  population of $N$ individuals and their concurrent social contacts within a given geographical area (a ``country").  In this study, we take that area to be a square of sides $L=a\sqrt{N}$ (with periodic boundary conditions), for the sake of simplicity.  The scaling of $L$ with $\sqrt{N}$ is selected so as to attain a specified global population density, $\rho_{\rm global}=N/L^2=1/a^2$.  Given the population density of the country one wishes to simulate, there is not really much choice regarding $N$ and $a$ that ought to be scaled, in simulations, to fit $\rho_{\rm global}$ (one could even use the actual shape of the country in question, instead of a square).

To account for a scale-free degree distribution of the network of social contacts, we employ a variant of the KR network growth redirection model~\cite{KR}: Each time we add a new node $y$ to the  network, we choose one of the existing nodes, $x$, at random and connect $y$ to $x$ with probability $(1-r)$; otherwise, with probability $r$, we connect $y$ to a neighbor of $x$, $x'$ (one of the nodes linked to $x$ by just one link), selected at random~\cite{kr_remark}.  This results in a scale-free network of degree exponent $\g$ close to the Krapivsky-Redner result of $\g_{KR}=1+1/r$.  There are not many data available for $\g$, certainly not at the scale of a whole country. We use $\g\gtrsim2$, typical of networks of friends on {\it facebook}, email networks, and similar social networks, which might serve as a proxy~\cite{muchnik}.  For most of the simulations presented below, we set $r=0.8$ ($\g\approx2.3$), noting, however, that $r$  is truly an adjustable parameter of the model.

While many other network growth techniques result in scale-free distributions, the KR model is particularly well suited for social networks of contacts, as it mimics key events of real-life networking: Some of our acquaintances we acquire directly, while many other contacts are established through referral (redirection) from old acquaintances, work, etc.  One thus hopes that the KR model captures not only the characteristic scale-free distribution but also important correlations between the degrees of the nodes.

The final aspect of the model pertains to how to place each node in space.  The first node is placed at the center of the square, or at any other arbitrary  location.  The placement of any subsequent new node $y$ depends on whether it joins the network directly (by connecting to a random $x$) or is redirected (to a random $x'$ neighbor of $x$).  A node that joins directly is placed at $(s, \theta)$ from $x$ (using polar coordinates), where the angle $\theta$ is chosen at random, and uniformly, from $0\leq\theta<2\pi$, and the distance $s$ is taken from the distribution $p(s)=Cs^{-\a};$ $b<s<S$, where $C$ is an appropriate normalization constant. For our square country, we use the cutoff distance $S=\sqrt{2}L$ so as to cover the whole region (recall that we also employ periodic boundary conditions). If $y$ joins, by redirection, to $x'$, then we simply place it at a distance $(b,\theta)$ from $x'$, at a random $\theta$. The distance $b$ is then the closest a new node can be placed to the node it connects to.  Since our fundamental unit of length is already fixed by $a=1/\sqrt{\rho_{\rm global}}$, we need to specify the ratio $b/a$.  For the present study, we set $b/a=1$: Using current data for the population density, this choice corresponds to $b=a\approx63\,$m for the United Kingdom and $b=a\approx180\,$m for the USA.  Though reasonable, our choice of $b/a=1$ is arbitrary, and this ratio is another free parameter of the model.

\begin{figure}[h]
\includegraphics[width=0.45\textwidth]{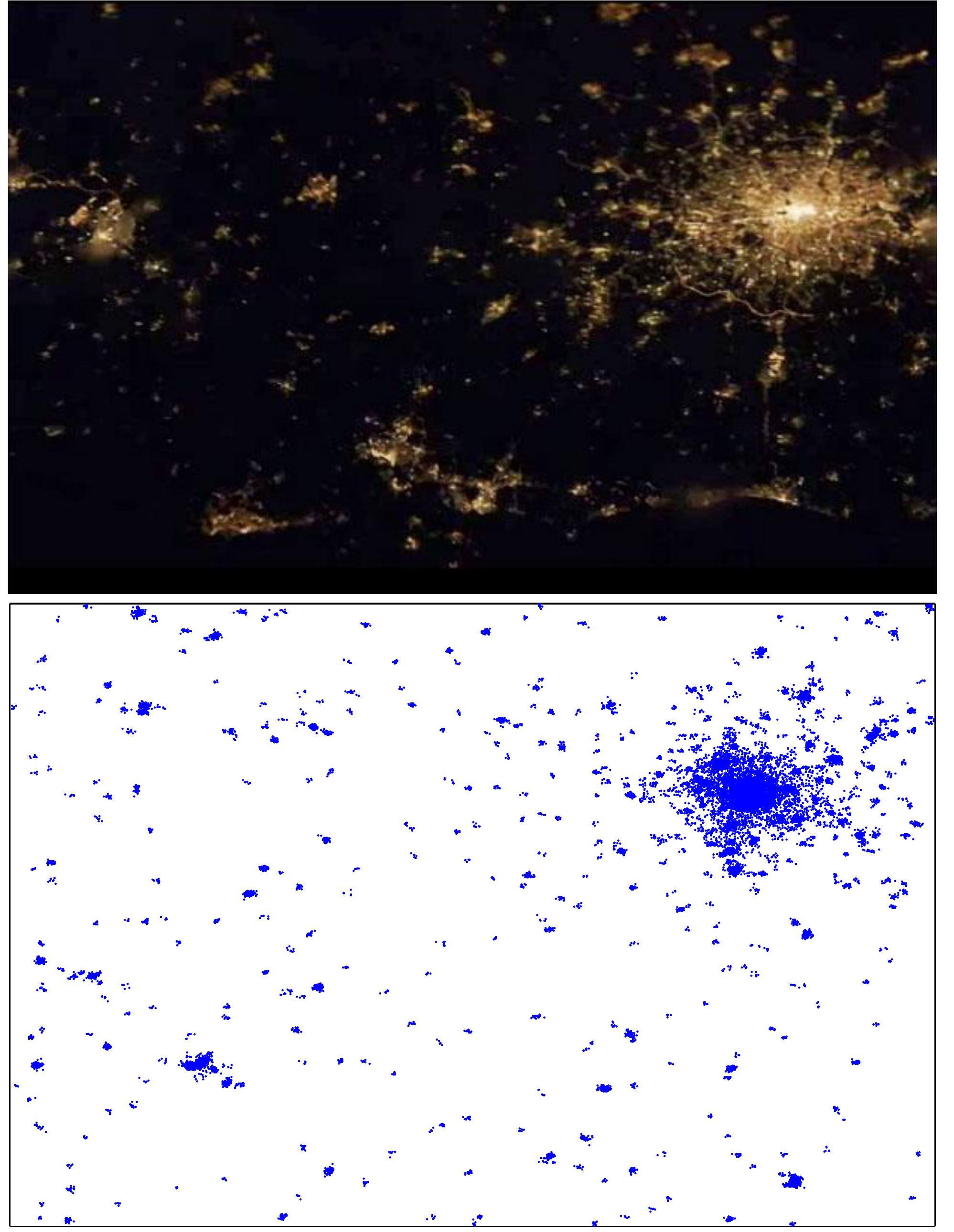}
\caption{Satellite picture of south England at night~\cite{sat} (top panel) and computer simulation snapshot (bottom panel), selected in this case among similar results because of its particularly close resemblance to reality.  The grid used for the simulation consists of $M=100$ divisions, corresponding to $n_{\rm th}=5$, and underoccupied boxes, with $n<n_{\rm th}$, have been filtered out.}
\label{south_england.fig}
\end{figure}

The idea behind choosing a power-law distributed distance comes from the seminal work of Kleinberg~\cite{kleinberg}, which shows that $\a=1$ is the optimal exponent for finding minimal paths between nodes in a two-dimensional lattice, based on {\it any} decentralized algorithm.  Finding such paths is important, for example, in Milgram's famous small-world experiment~\cite{milgram}, and might therefore be relevant for the cohesion and proper function of society as a whole.  In addition, we tried several values for the distance exponent $\a$ and found that the spread of nodes best resembles satellite pictures of the Earth at night when $\a$ is close to 1 (see Fig.~\ref{south_england.fig}).  

To summarize, this simple stochastic model relies on a handful of tunable parameters: the redirection probability $r$ (or degree exponent $\gamma\approx1+1/r$), the distance exponent $\alpha$, and the ratio $b/a$ (there is really little choice regarding $L$ and $N$, which are determined by the desired population density and the computer's limitations).  As a first, naive attempt at setting these parameters, we have relied on the typical connectivity observed in nets in a social context, on Kleinberg's work, and on the resemblance of the model's results to satellite pictures of the Earth at night.  The comparison to satellite pictures was also useful in designing elementary aspects of the model.  For example, it has guided the search for an appropriate adaptation of the KR net growth model~\cite{kr_remark}, and we thus found that choosing the same power-law distance for nodes that join the net directly or by redirection  yields a poorer resemblance than the actual split rule we have settled upon.  

\section{analysis and results}
\label{results.sec}
For our simulations, we use $N=50,\!000$, $r=0.8$,  $\alpha=1$, and $b/a=1$,  and a square country of sides $L=a\sqrt{N}$.  In order to distinguish between the various ``cities,"  we use the City Clustering Algorithm (CCA),  proposed by Rozenfeld {\it et al.}~\cite{hernan}: The country is superposed by an $M\times M$ square grid, subdividing it into square boxes of sides $\ell=L/M$. A box is deemed {\it occupied} if the number of nodes  within it, $n$, exceeds the threshold $n_{\rm th}=\rho_{\rm global}\ell^2=(N/L^2)\ell^2=N/M^2$, the average number of nodes expected in each box in a random {\it uniform} scattering of the nodes.  Clusters of occupied cells define the various cities; two adjacent occupied cells belong to the same cluster (or city) if they share an edge.  As in Ref.~\cite{hernan}, we find that the results show little variation for a range of $M$ (or $\ell$), so long as $n_{\rm th}$ is small, of the order of 1. However, for the growth rate of cities, we observe a qualitative shift when the boxes become big enough (see Sec.~\ref{results.sec}\,{\bf C}). For concreteness, we use $M=50$, $100$, and $225$, corresponding to $n_{\rm th}=20$, $5$, and~$1$.

\subsection{Theory: The limit of $b/a\to0$}
A theory for the model can be constructed by imagining that the country is filled with $N$ nodes, one step at a time.  Let the number of nodes in ``city" $i$ after $t$ steps be $n_i(t)$, and let the total number of links coming out of the nodes in the city be $m_i(t)$.  These quantities are constrained by an overall normalization: 
\begin{equation}
\label{norm.eq}
\sum_i n_i(t)=t,\qquad \sum_im_i(t)=2t\,, 
\end{equation}
since exactly one node and two links are introduced to the system {\it as a whole} in each step (the link is counted twice, once for each node it connects to).

We can write exact evolution equations for $n_i(t)$ and $m_i(t)$, valid for the limiting case of $b/a\to0$. In this limit, the only way for  a node to join  city $i$ is by attaching to one of the city nodes {\it by redirection}, so that it lands within distance $b$ from that node (and we assume $\ell>b$). Nodes that attach to a city node {\it directly} land almost surely outside of the city, since the range $(b,L)$ is infinite, but they do add a link to $m_i$.  The limit $b/a\to0$  guarantees that nodes connecting to {\it other} cities land almost surely outside of city $i$, leaving it unaltered.  Finally, the probability of two cities coming into contact and merging tends to zero, in this limit, further simplifying the analysis. To summarize, the processes affecting $n_i(t)$ and $m_i(t)$ during the next step are
\begin{equation}
\label{nmupdate.eq}
(n_i,m_i)\longrightarrow
\begin{cases}
(n_i+1,m_i+2) &\mbox{prob: }r\frac{m_i}{2t}\,,\\
(n_i,m_i+1) &\mbox{prob: }(1-r)\frac{n_i}{t}\,,
\end{cases}
\end{equation}
where the first line is the case of a node joining by redirection (note that the city gains two links, in addition to the node itself), and the second line denotes the case where a city node is selected to connect to it directly.  For each of these events, we have used the normalization relations~(\ref{norm.eq}).

With the assumptions above, we have
\begin{eqnarray}
\frac{d}{dt}n_i(t)&=&r\frac{m_i(t)}{2t}\,,\label{n.eq}\\
\frac{d}{dt}m_i(t)&=&(1-r)\frac{n_i(t)}{t}+2r\frac{m_i(t)}{2t}\,,\label{m.eq}
\end{eqnarray}
where $n_i$ and $m_i$ here express average (expected) quantities. These equations assume that $n_i$, $m_i$, and $t$ are  {\it continuous} variables and are therefore not valid for $t\approx t_i$, the time when the first city node was introduced (they ignore fluctuations in number space). They are meaningful, however, in the long-time asymptotic limit, whose solution is
\begin{equation}
\label{nmsol.eq}
n_i(t)=\left(\frac{t}{t_i}\right)^{\beta},\qquad m_i(t)=\kappa n_i(t)\,,
\end{equation}
where $\beta=\frac{r}{2}+\frac{1}{2}\sqrt{r(2-r)}$ and $\kappa=1+\sqrt{(2-r)/r}$.
The constant $\kappa$ can be understood as the average connectivity per node ($m_i/n_i$) for large cities, or for $t/t_i\gg1$.

The limit $b/a\to0$ corresponds to the unrealistic case of pointlike cities (or alternatively, cities of finite diameter, but infinitely far apart from one another).  Nevertheless, this simple limit captures a lot of the qualitative behavior displayed by the model.

\subsection{Zipf's law}
One can rank the cities of any given country according to their population sizes: The largest city is number 1, the second largest is number 2, etc.  The populations of the cities typically decay inversely proportional to their rank~\cite{zipf,Eeckhout,levy,holmes}.  This is known as Zipf's law (originally proposed by Zipf for word frequency, by rank).

The  distribution of city populations can be obtained from Eq.~(\ref{nmsol.eq}) as follows~\cite{ptrick}. 
Since new cities are seeded every time a node joins the network directly, and that happens at a constant rate, the probability that city $i$ has been seeded until time $T$ (or that $t_i<T<N$) is $T/N$.  Using $t_i=T$, $t=N$, in Eq.~(\ref{nmsol.eq}), and inverting the equation, we obtain   
$T(n_i)=Nn_i^{-1/\beta}$.
Then, the city population distribution is given by
\begin{equation}
p(n_i)=-\frac{\partial T(n_i)}{\partial n_i}=\frac{N}{\beta}n_i^{-1-1/\beta}\,.
\end{equation}
This implies a power-law Zipf-like relation for the tail of the population of cities by rank $s$, of the form (see, for example, Ref.~\cite{cbp})
\begin{equation}
\label{ns.eq}
n_s\sim s^{-\beta}\sim s^{-\frac{r}{2}-\frac{1}{2}\sqrt{r(2-r)}}\,.
\end{equation}

\begin{figure}[h]
\includegraphics[width=0.45\textwidth]{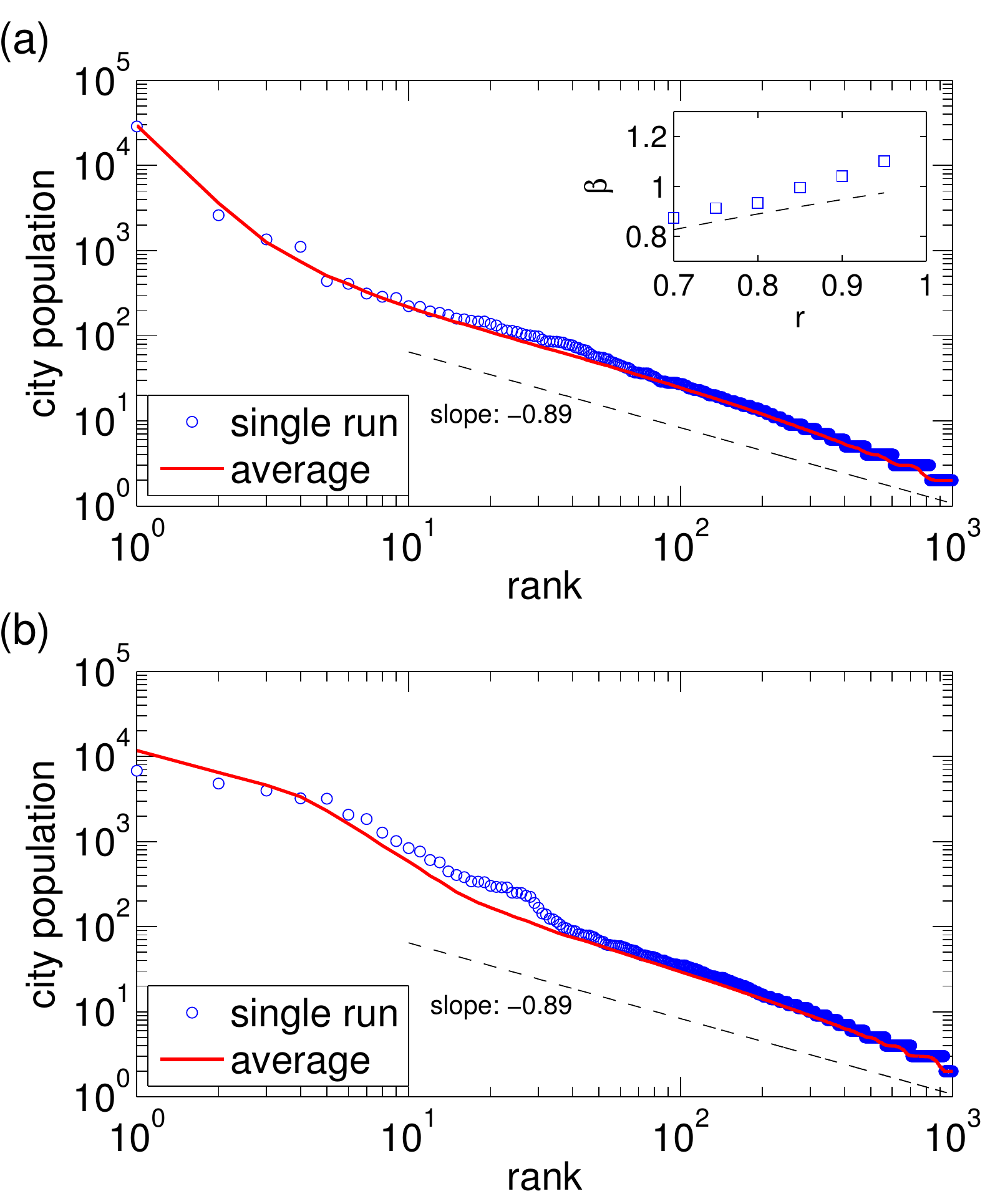}
\caption{Rank-population plot for a single configuration and for the proper average over 100 runs when the network starts with two nodes~(a) and when the network starts with ten interconnected nodes uniformly spread in the spatial domain~(b) (see Discussion). Dashed lines indicate the predicted slope of the tail for $r=0.8$ of the simulations shown. The fitted slope of the tail of the rank-population curve (for cities of rank $\geq 10$) is $-0.930$, with $R^2>0.94$ and the $95\%$ confidence interval of $[-0.938,-0.922]$, for panel (a). The inset in~(a) shows the measured exponent as a function of $r$ and is compared to the theoretical prediction of Eq.~(\ref{ns.eq}). Note, in particular, that the Zipf exponent $-1$ can also be achieved for some value of $r$ ($\approx 0.85$). The grid used for the CCA algorithm has $M=225$ divisions, corresponding to $n_{\rm th}\approx1$.}
\label{Zipf.fig}
\end{figure}

A big advantage of the model is that one can simulate a specific scenario a large number of times, yielding high quality data---an option that is simply not available for real data, since the number of real countries is limited, and their specific conditions might differ markedly, making their comparison moot.  To illustrate this point, we ran 100 simulations of the same parameters and averaged over the populations of cities of corresponding ranks, across the different runs (for a possible caveat, see Ref.~\cite{cbp}).  The average populations plotted against their rank, as well as the noisier data from a single run, are shown in Fig.~\ref{Zipf.fig}. The power-law exponent measured for the tail from the average is compared to $\beta$ of Eq.~(\ref{ns.eq}), for a range of $r$ values, in the inset of the figure.  The differences, which are larger as $r\to1$, can be ascribed to the neglected effects in the $b/a\to0$ limit. The largest cities average to quite larger than the background power law.  This is explained by merging, which provides the largest cities with a faster way of growth, way beyond  the basic node-by-node accumulation described in Eqs.~(\ref{n.eq}) and (\ref{m.eq}).

\subsection{Gibrat's law}
We now examine the rate of city growth.  Suppose that in a certain period of time, $dt$, a city of population $n$
grows to $n+dn$.  The {\it growth factor} of the city (for that time period) is then given by~\cite{hernan}
\begin{equation}
g(n)=\ln\left(\frac{n+dn}{n}\right)\approx\frac{dn}{n}\,.
\end{equation}
Using Eq.~(\ref{n.eq}), we get
\begin{equation}
\frac{dn}{n}=\frac{r}{2}\frac{m}{n}\frac{dt}{t}\,.
\end{equation}
Thus, the growth factor of a city depends on the ratio $m/n$.  For large cities, the ratio $m/n\to\kappa$, and we expect a constant growth factor (independent of $n$).  If cities $n_1$ and $n_2$ merge, during the period $dt$, then we define the growth factor for the merged city as $g(n_1+n_2)=\ln[(n_1+dn_1+n_2+dn_2)/(n_1+n_2)]\approx(dn_1+dn_2)/(n_1+n_2)$, after Ref.~\cite{hernan}. In that case, if either $n_1$ or $n_2$ is large, then the growth factor of the merged city tends to the same constant  as for large cities, $g=r\kappa dt/2t$. In other words, the growth factor  for large cities is constant, even if merging events are included (they can be avoided, formally, by making $dt$ small enough).  Gibrat's law states that both $g(n)$ and its variance, $\sigma_g(n)\equiv\sqrt{\av{g(n)^2}-\av{g(n)}^2}$, are independent of $n$.

The ratio $m_i/n_i$ and its distribution can be studied from the master equation for $N_{nm}(t)$,  the {\it number} of cities of $n$ nodes and $m$ links at step $t$, constructed from the processes in Eq.~(\ref{nmupdate.eq}):
\begin{equation}
\label{Nnm.eq}
\begin{split}
&N_{nm}(t+1)-N_{nm}(t)=(1-r)\frac{n}{t}[N_{n,m-1}(t)-N_{nm}(t)]\\
&+\frac{r}{2t}[(m-2)N_{n-1,m-2}(t)-mN_{nm}(t)]
+(1-r)\delta_{n,1}\delta_{m,1}\,.
\end{split}
\end{equation}
This equation can be analyzed by the van Kampen's $1/\Omega$ expansion; otherwise, the time variable can be scaled away using the ansatz $N_{nm}(t)\sim f_{nm}t$, and the resulting master equation for the $f_{nm}$'s can be studied by standard techniques~\cite{vankampen}.
Either way, one finds that $\av{m/n}$ tends to a constant, as well as a variance of $\av{(\frac{m}{n})^2}-\av{\frac{m}{n}}^2\sim1/n$, thus predicting that $\sigma_g(n)\sim1/\sqrt{n}$.  

\begin{figure}[h]
\includegraphics[width=0.45\textwidth]{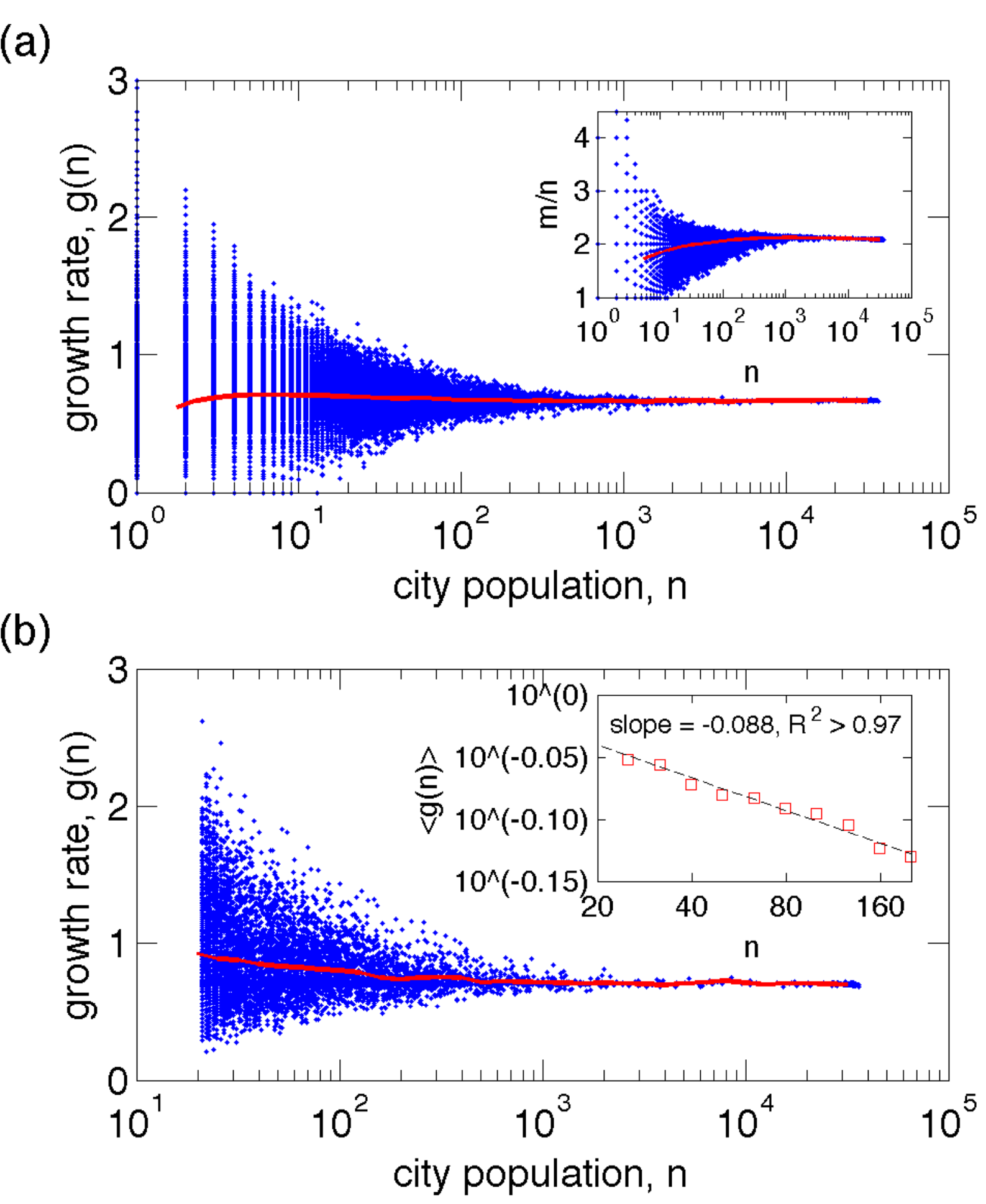}
\caption{Growth factor by population size, for $M=225$~(a) and $M=50$~(b).  The inset in~(a) shows the scatter plot of $m_i/n_i$, by population size.  Note the resemblance to the distribution of $g(n)$.  The average growth factor $\av{g(n)}$ is shown as a solid line in both panels.  Note the subtle shift as the box size increases: For the larger boxes in~(b), there appears an initial weak power-law decay, consistent with an exponent 
$x$ in the $95\%$ confidence interval: [0.078,0.098] (inset).}
\label{gibrat.fig}
\end{figure}

\begin{figure}[h]
\includegraphics[width=0.45\textwidth]{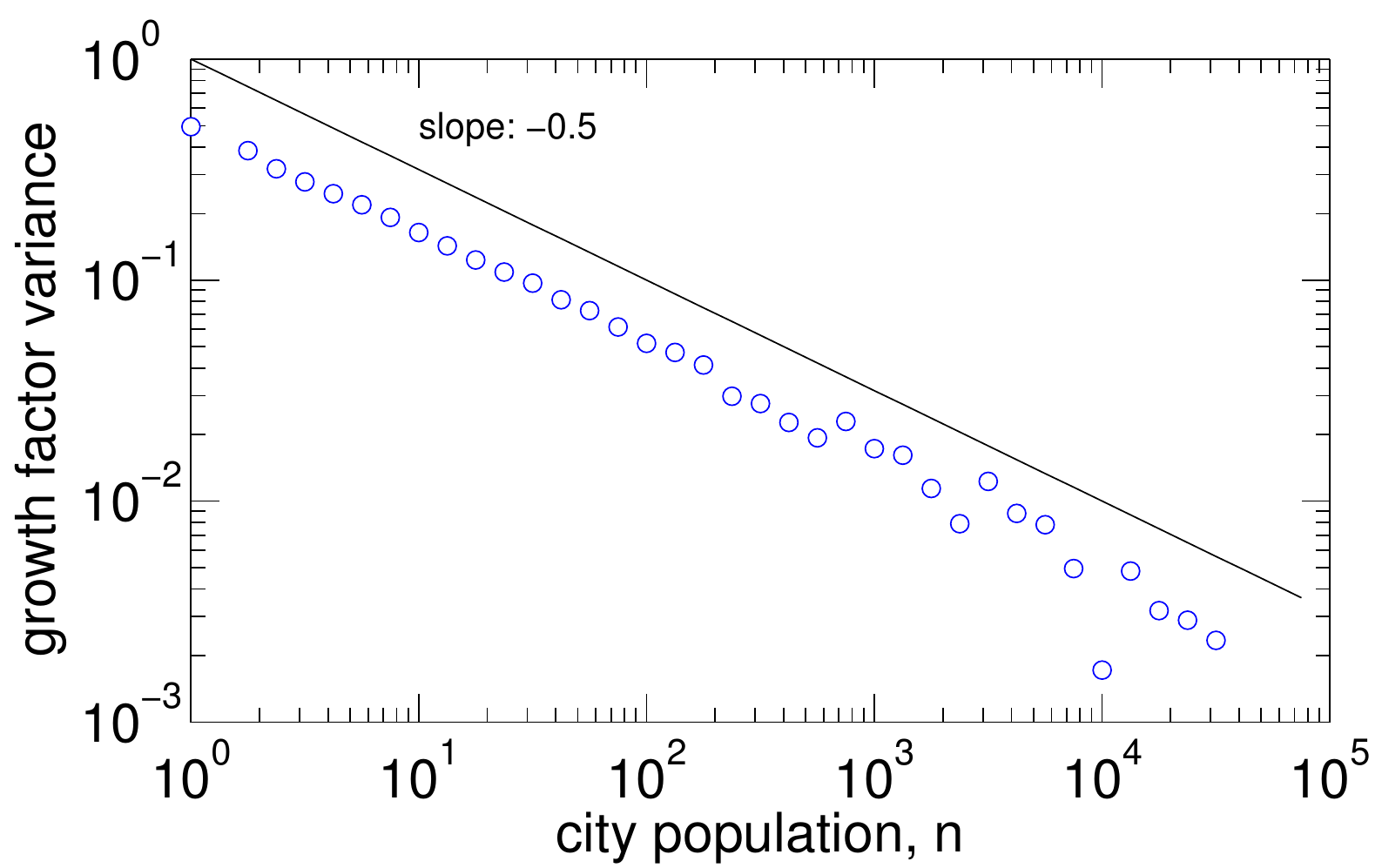}
\caption{Growth factor variance by population size, for $M=225$.  The solid line represents a slope of $-1/2$, which is expected from the theory for the $b/a\to0$ limit.}
\label{gibrat_sig.fig}
\end{figure}

In Fig.~\ref{gibrat.fig}, we show the scatter of $g(n)$ as a function of city population $n$.  The inset of panel~(a) shows the spread of $m_i/n_i$, for the same simulation parameters, demonstrating the close resemblance between the two quantities.  The average growth factor $\av{g(n)}$ is nearly constant, but subtle differences appear as the box size grows from $M=225$ in panel~(a) to $M=50$ in panel~(b).  The decay of $\sigma_g(n)$ in simulations,  shown in Fig.~\ref{gibrat_sig.fig}, is in reasonable agreement with the predicted $1/\sqrt{n}$ decay (there is little difference in this result as the box size increases).

It is tempting to compare the simulation results to real-life data, despite the crudeness of our model.  In Ref.~\cite{hernan}, it was found that growth factors and their variance exhibit a power-law decay, $\av{g(n)}\sim n^{-x}$ and $\sigma_g(n)\sim n^{-y}$, with $x\approx0.17, 0.21, 0.28$ and $y\approx0.27, 0.19, 0.20$ for Great Britain, Africa, and the USA, respectively.   Our model agrees with these findings in that the variance decays as a power law, though the exponent $y\approx0.50$ is quite a bit larger than for real data.  The weak power-law decay found for the growth factor itself in the real data
is observed in our model only when the grid box is large enough~[Fig.~\ref{gibrat.fig}b, inset] and only for cities up to a certain size: The growth factor of large cities is flat, in agreement with Gibrat's law.

\subsection{Population density vs.~city population size}
We next turn to the population density $\rho_i$ as a function of city population $n_i$. Consider, first, a small city $i$ that perhaps fits within a single $\ell\times\ell$ box.  If $b<\ell$, then during the first stage of the city's growth, its population might increase while still fitting within that single box.  During that initial phase, the density would grow linearly with $n_i$: $\rho_i=n_i/\ell^2$.  Eventually, the city would spill into an adjacent box, resulting in an immediate twofold density reduction, and the density would resume linear growth with $n_i$, until it spills into the next box, etc.  

For larger cities, occupying clusters of several boxes, we can ignore the discrete geometry of the boxes and estimate the density growth from the basic theory for the limit $b/a\to0$.  Recalling that a city then grows only when a node joins it by redirection (and settles within distance $b$ of the target node), the diameter $D(n)$, of a city of population $n$, grows only when the target node is at the city's rim.  A node from the boundary, as opposed to an interior node, is selected as a target with a probability proportional to $1/\sqrt{n}$.  When selected, the diameter $D(n)$ grows proportionally to $b/\sqrt{n}$, since there are of the order of $\sqrt{n}$ nodes on the rim.  Thus, 
\begin{equation}
\frac{dD(n)}{dn}\propto\frac{1}{\sqrt{n}}\frac{b}{\sqrt{n}}\,,
\end{equation}
leading to $D(n)\sim\ln n$ and $\rho(n)\sim n/\ln^2n$.

\begin{figure}[h]
\includegraphics[width=0.45\textwidth]{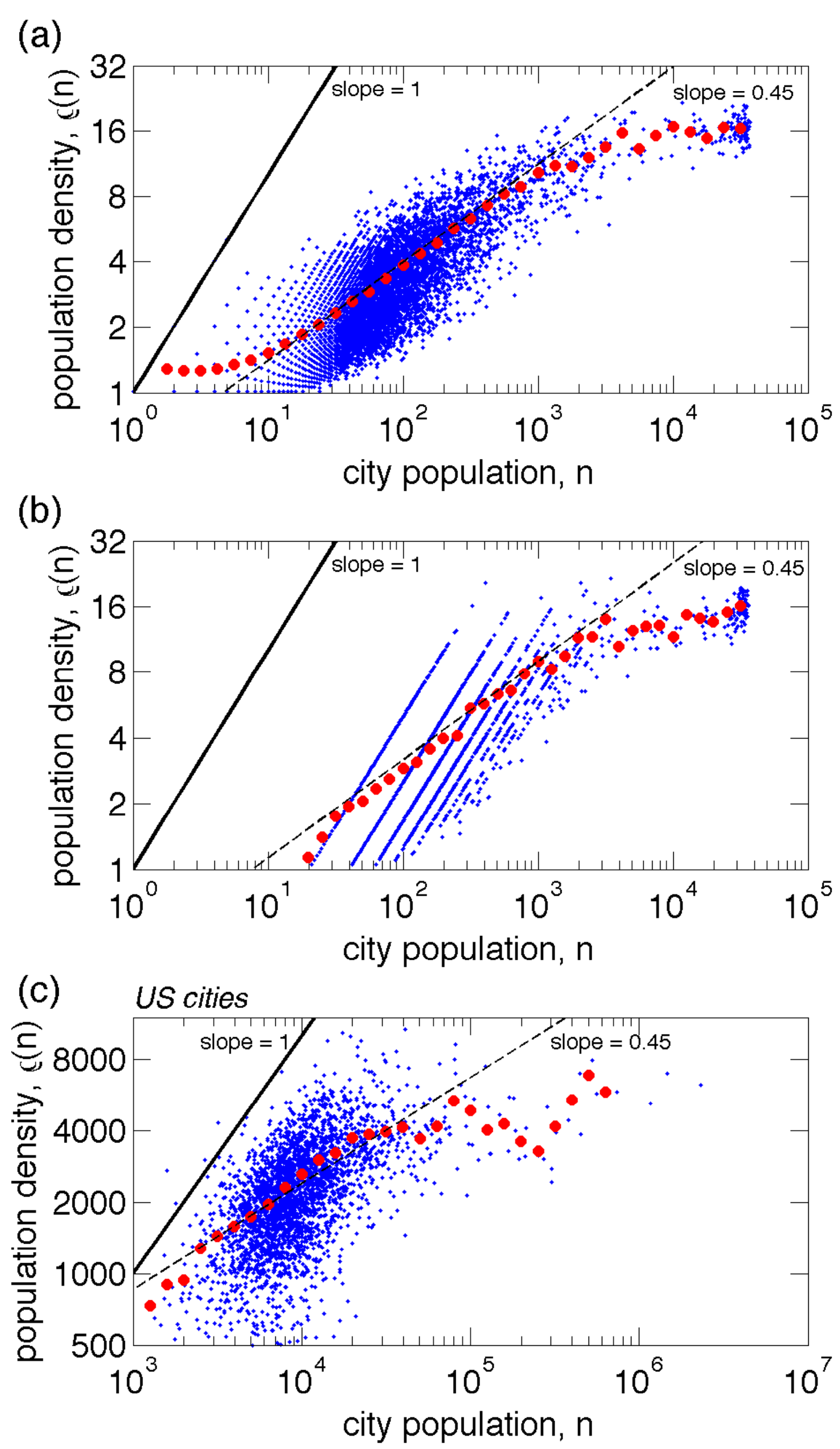}
\caption{Population density by city population size, for (a)~$M=225$ and (b)~$M=50$, and (c)~real data for the USA, from Ref.~\cite{hernan}. The solid lines represent the expected linear bound, $\rho\sim n$, due to the boxes' discreteness, which is more apparent for the larger boxes in panel~(b).  An effective slope of $\sim0.45$ (dashed lines) is consistent with the averages ($\bullet$) of all three data sets.}
\label{density.fig}
\end{figure}

In Fig.~\ref{density.fig}, we show simulation results for the population density as a function of city population size. The linear growth predicted for small cities due to the discrete effect of the boxes is more pronounced the coarser the grid, and it is clearly visible in panel~(b), for $M=50$.  A wide scattering of $\rho(n)$ is observed, both in simulations, (a) and (b), and real data~(c), excluding any meaningful comparison to the prediction that $\rho(n)\sim n/\ln^2n$.  Interestingly, real data for the USA~\cite{hernan}, in panel (c), show a very similar scatter to that of the theoretical model: The average density per population size seems consistent with an effective power law, $\av{\rho(n)}\sim n^{0.45}$, and $\av{\rho(n)}$ starts to slowly saturate as the city population increases.

We ascribe the saturation of $\av{\rho(n)}$ in the model to merging effects: As large cities merge, their density is dampened because of the large increase in area.  For real life data, there might be an additional reason, not mirrored in our model, dictated by architectural constraints and limits of comfort.  A similar scatter, and the effective power-law growth $\av{\rho(n)}\sim n^z$, is observed also in data for the world's 400 largest cities, gathered from the Internet~\cite{citymayors}, but with exponent $z\approx0.83$.  Most probably, though, the effective power law has no deep underlying basis and is simply caused by the scattering of the data.   For further analysis of $\rho(n)$, see Ref.~\cite{hernanb}.

\subsection{Social connectivity}
\begin{figure}[]
\includegraphics[width=0.45\textwidth]{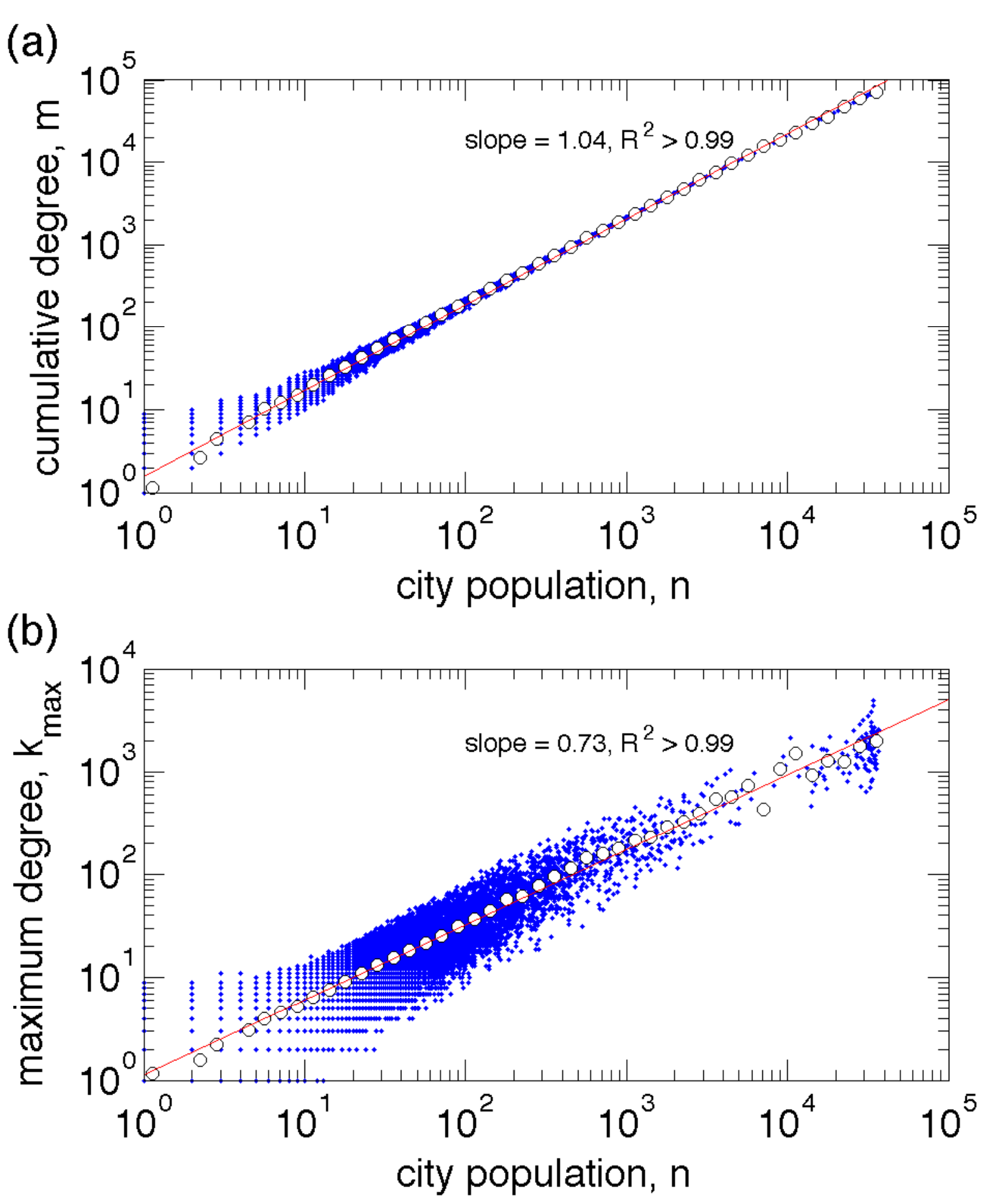}
\caption{Scatter plot of: (a) the cumulative degree of all nodes, and (b) the degree of the most connected nodes as a function of the city size, respectively.
In (a) the average cumulative degree ($\circ$) follows a superlinear power-law scaling (solid line), $\langle m(n)\rangle\sim n^{y}$, with an estimated exponent in $[1.03,1.05]$  ($95\%$ confidence interval).
In (b) the average maximum degree ($\circ$) follows a sublinear power-law scaling (solid line), with an estimated exponent in $[0.72,0.75]$  ($95\%$ confidence interval), consistent with the theoretical prediction of $\av{k_{\rm max}(n)}\sim n^{1/(\g-1)}$, where $\g\approx2.3$ for $r=0.8$.
The leveling off of the curve towards larger $n$ is explained by the effect of merging events.
}
\label{connectivity.fig}
\end{figure}

The model has been constructed in such a way that the social network of connections is scale-free, with degree exponent $\gamma\approx1+1/r$. Our zeroth-order theory also predicts a linear relation between the {cumulative degree} of a city population, $m_i$, and the city size, $n_i$, Eq.~(\ref{nmsol.eq}). The actual data, however, plotted in Fig.~\ref{connectivity.fig}(a), exhibit a weak superlinear relation, $m_i\sim n_i^y$, with an estimated $y$ exponent in the interval of $[1.03,1.05]$ with $95\%$ confidence.  Intriguingly, a similar superlinear correlation was  found recently for real data, in the ingenuous study of Schl\"{a}pfer {\it et al.}~\cite{Schlapfer2012}.  The superlinear growth in our model can only be ascribed to correction effects, too subtle to be captured by our coarse theory.  The resemblance to real data is encouraging nevertheless.

We were also hoping to capture the notion that highly connected individuals tend to live in more populated areas.
Simulations of the model seem to confirm this idea.  In Fig.~\ref{connectivity.fig}, we show a scatter plot of the degree of the most connected node in city $i$, as a function of $n_i$.  The scaling of the maximum degree as a power of $n_i$ can be understood as follows.
Assume that the $n_i$ nodes in city $i$ form a random (uncorrelated) subset of the total number of nodes in the system.  In that case, the degree distribution of the city nodes is the same as the degree distribution of all nodes, that is, $p(k)\sim k^{-\g}$.  Given $n_i$ nodes obeying such a distribution, the maximum degree in the set scales
as $k_{\rm max}(n_i)\sim n_i^{1/(\g-1)}$.  Indeed, $\av{k_{\rm max}(n_i)}$ agrees quite nicely with this prediction, especially for small cities, suggesting that correlations  are not too important in this context.  Large cities, on the other hand, undergo merging events so that their total number of nodes is actually the sum of several {\it independent} distributions (each consisting of a smaller number of nodes).  Thus, the maximum degree in large cities is effectively smaller than the predicted $n_i^r$, explaining the leveling off of the curve towards large $n_i$.

\section{Discussion}
\label{discussion.sec}

In conclusion, we have presented a simple stochastic process to model the geographical spread as well as the network of social connections in human populations.  In order to increase transparency, we intentionally kept the number of free parameters at a minimum.  Despite the model's simplicity, it successfully reproduces major aspects of real-life populations, including a Zipf-like ranking of cities by population size
(but for the largest cities), a scale-free network of social contacts, and a superlinear correlation between social connectivity and city population.

There are not many known results for the social network of connectivities, at the scale of whole countries.  Nevertheless, the main expected features, that the degree distribution of contacts is scale-free and that highly connected individuals live in the more populated areas, are well reflected in our model.  The least realistic feature of our network of connections is that it is a {\it tree}, as dictated by the KR algorithm used for its production.  Real social networks are characterized by a high level of {\it clustering}, where nodes that share a neighbor tend to be connected to one another (your friends are likely to be friends of one another).  A simple solution to this problem 
is to connect each of the existing nodes to a certain (fixed or random) number of nodes, closest to it geographically.  This would reflect the fact that one is usually acquainted with a certain number of neighbors.  Such a procedure would add a constant number of connections to each node's degree and would not affect the overall scale-free distribution, while creating numerous loops and triangles, bolstering the clustering of the network. 

The model is simple enough to allow for some analysis.  We have presented various derivations, valid for the limit of $b/a\to0$.  Deviations from this limit might be treated as perturbations, though their effect might be pronounced.
Besides merging, which affects mostly the biggest cities in ways discussed throughout the paper, the largest correction to the $b/a\to0$ limit comes from the event that a new node connects to a city node {\it directly} but lands inside the city rather than landing outside (as guaranteed by the $b/a\to0$ limit).  In this case, the city gains a node and two links, which is the same as for a redirected connection [line 1 of Eq.~(\ref{nmupdate.eq})].  The probability for this event could become pretty large, for $\alpha=1$, despite the fact that the city diameter is much smaller than $L$.

Of the  many obvious ways in which one could improve the model (and that we have resisted, to avoid complications), perhaps the most significant would be to allow for a small fraction of the population to migrate from city to city.  Even a tiny fraction of translocations would have a profound effect on the network of contacts, since when  a node moves from city $i$ to $j$, it retains its connections with other nodes at $i$.
Another possible extension to the model is to start the growth process with a random network of ``seed" nodes distributed uniformly in space, resembling the spatial diversity of large cities seen in real life. This will overcome one of the unrealistic aspects of the current model, which tends to produce mega-cities that are oversized. Our preliminary results presented in Fig.~\ref{Zipf.fig}(b) show some promise and give us hope that similar manipulations could reproduce the Zipf law observed for the largest cities.

Exploring the relation between the network of connections and the spatial distribution of nodes had been one of our primary goals in embarking on the present study.  Since real-life data for large populations, their locations, {\it and} the concurrent  network of contacts are not readily available, the promise of our model is in that it provides a plausible, simple basis for the study of various questions of social dynamics, such as how epidemics, or rumors and opinions would evolve, and how these would be affected by the model's (few) parameters. We could now, for example, simulate Milgram's experiment and try to interpret the results in the light of Kleinberg's work but in a far more realistic---yet manageable---setup than the original small-world square lattice employed in his seminal work.

\acknowledgements
We thank James Bagrow for his involvement in earlier stages of this work, and we thank Erik Bollt for many useful discussions.

\end{document}